\documentclass{nlaauth}
\usepackage{graphicx,wrapfig,bm}
\usepackage{pifont,latexsym,ifthen,theorem,rotating,calc}
\usepackage{amsfonts,amssymb,amsbsy,amsmath}
\newtheorem{prop}{Proposition}[section]
\begin{document}

\NLA{1}{16}{xxx}{yyy}{07}
\runningheads{G.~Ito}{Least change in the Determinant or Permanent}
\noreceived{}
\norevised{}
\noaccepted{}
\title{Least change in the Determinant or Permanent of a matrix under perturbation of a single element: continuous and discrete cases}
\author{Genta~Ito\corrauth}
\address{Maruo Lab., 500 El Camino Real \#302, Burlingame, CA 94010, United States.}
\corraddr{Maruo Lab., 500 El Camino Real \#302, Burlingame, CA 94010, United States. (Email: gito@maruolab.com)}
\begin{abstract}
We formulate the problem of finding the probability that the determinant of a matrix undergoes the least change upon perturbation of one of its elements, provided that most or all of the elements of the matrix are chosen at random and that the randomly chosen elements have a fixed probability of being non-zero.  Also, we show that the procedure for finding the probability that the determinant undergoes the least change depends on whether the random variables for the matrix elements are continuous or discrete. 
\end{abstract}
\section{Introduction}\label{SecIntro_p2}
Consider the problem of finding the probability that the determinant of a matrix undergoes the least possible change upon perturbation of one of its elements. 
In this paper, we present non-asymptotic solutions for that class of matrices and others, and we discuss the connection between the case where the (real) random variables for the matrix elements are continuous and the case where they are discrete.
Denote the determinant of any matrix $A$ by $\det A$. For an $(n+1)\times(n+1)$ matrix $M_{n+1}$, the expansion of $\mathrm{det}$\,$M_{n+1}$ via row $i$ is
\begin{equation}\label{DefDeterminant1_p2}
\mathrm{det}\, M_{n+1}=m_{i1}M_{i1}+\cdots+m_{ij}M_{ij}+\cdots+m_{in+1}M_{in+1},
\end{equation}
where $m_{ij}$ is the element of $M_{n+1}$ at the intersection of row $i$ and column $j$, and $M_{ij}$ is the cofactor of $m_{ij}$. Since $m_{ij}M_{ij}$ is the only term in that expansion of $\det M_{n+1}$ in which element $m_{ij}$ occurs, the probability that perturbation of $m_{ij}$ has the least effect on the value of $\mathrm{det}$\,$M_{n+1}$ is equal to the probability that $|M_{ij}|$ is as small as possible.
The cofactor $M_{ij}$ of element $m_{ij}$ in $M_{n+1}$ is $(-1)^{i+j}\det S_{n}$, where $S_{n}$ is the $n\times n$ submatrix of $M_{n+1}$ which is obtained by deleting row $i$ and column $j$. Thus $|M_{ij}|=|\det S_{n}|$, so the probability that $\det M_{n+1}$ undergoes the least change upon perturbation of element $m_{ij}$ is equal to the probability that $|\det S_{n}|$ is as small as possible. In particular, perturbation of $m_{ij}$ has no effect at all on $\det M_{n+1}$ if $\det S_{n}=0$.
In this paper, we treat the following three classes of $n\times n$ matrices $S_{n}$:
\begin{enumerate}
\item[(i)]Matrices $A_{n}$ in which all the elements are values of mutually independent random variables each of which has probability $r$ of being non-zero and probability $1-r$ of being 0, where $0<r<1$.
\item[(ii)]Matrices $B_{n}$ in which all but one of the diagonal elements are set to 1 (i.e., $b_{ii}=1$ for $i\ne 1$, where $b_{11}$ is the special diagonal element), and $b_{11}$ and all the off-diagonal elements are as in (i). Thus  
\begin{equation}\nonumber
B_{n}=\left(\begin{array}{ccccc}b_{11}&b_{12}&b_{13}&\cdots& b_{1n}\\
b_{21}&1&b_{23}&\cdots& b_{2n}\\
b_{31}&b_{32}&1&\cdots& b_{3n}\\
\vdots&\vdots&\vdots&\ddots&\vdots\\
b_{n1}&b_{n2}&b_{n3}&\cdots& 1\\\end{array}\right)
\end{equation}
\item[(iii)]Matrices $C_{n}$ in which all the diagonal elements are set to 1 and all the off-diagonal elements are as in (i).  Thus
\begin{equation}
\nonumber C_{n}=\left(\begin{array}{ccccc}1&c_{12}&c_{13}&\cdots&c_{1n}\\c_{21}&1&c_{23}&\cdots&c_{2n}\\c_{31}&c_{32}&1&\cdots&c_{3n}\\\vdots&\vdots&\vdots&\ddots&\vdots\\c_{n1}&c_{n2}&c_{n3}&\cdots& 1\\\end{array}\right)
\end{equation}
\end{enumerate}
The matrices $S_{n}$ described in (i), (ii), and (iii) above will be said to be matrices of type $\mathbf{A}_n,\mathbf{B}_n$, and $\mathbf{C}_n$, respectively. Every randomly chosen element of a matrix of any of these types (i.e., every element of a matrix of type $\mathbf{A}_n$, every off-diagonal element of a matrix of type $\mathbf{B}_n$ or $\mathbf{C}_n$, and the special diagonal element $b_{11}$ of a matrix of type $\mathbf{B}_n$) will be called a {\it variable} element; the remaining elements (those which are not chosen at random) will be called {\it fixed} elements.
Let $\mathfrak{X}$ be a set such that $0\in\mathfrak{X}$, let $X$ be a random variable whose values are in $\mathfrak{X}$, and 
let $P$ be the probability function for $\mathfrak{X}$ defined by
\begin{equation}\label{DefProbability_p2}
\begin{array}{rcl}
P(X=0)&=&1-r\\P(X\ne 0)&=&r\\
\end{array}
\end{equation}
If $D\subseteq\mathfrak{X}$ and $\mathfrak{X}$ is a discrete set of numbers, then $P(X\in D)$ can be expressed in terms of a probability function $p$ as
\begin{equation}\label{ProbDiscreteX_p2}
\mbox{$P(X\displaystyle \in D)=\sum_{x\in D}\,p(x)$,}\\
\end{equation}
where
\begin{equation}
\nonumber
0\leqq p(x)\leqq 1,\hspace*{.2in}p(0)=1-r,\hspace*{.2in}\mbox{$\displaystyle \sum_{x\in \mathfrak{X}-\{0\}}\,p(x)=r$}
\end{equation}
If $\mathfrak{X}$ is a continuous set of real numbers, then $P(X\in D)$ can be expressed in terms of a density function $q$ as 
\begin{equation}\label{ProbContinuousX_p2}
\mbox{$P(X\displaystyle \in D)=\int_{D}\,q(x)dx$,}
\end{equation}
where
\begin{equation}\nonumber
0\leqq q(x)\leqq 1,\hspace*{.2in}
\mbox{$\displaystyle\int_{\{0\}}q(x)dx=1-r$},\hspace*{.2in}\mbox{$\displaystyle \int_{\mathfrak{X}-\{0\}}\,q(x)dx=r$}
\end{equation}
In what follows, $q$ is assumed to be continuous on $\mathfrak{X}-\{0\}$. Then $\displaystyle \int_{D}q(x)dx=0$ if $D=\{x\}$ for some $x\in \mathfrak{X}-\{0\}$, which implies that 
\begin{equation}\label{FeatureProbOnContinuousX_p2}
P(X=x)=\left\{\begin{array}{rcl}1-r,&\hspace*{.2in}&x=0\\
0,&\hspace*{.2in}&x\in\mathfrak{X}-\{0\}\\\end{array}\right.
\end{equation}
\begin{prop}\label{prop1_p2}
Let $\mathfrak{X}$ be a continuous set of real numbers such that $0\in\mathfrak{X}$, and let $M_{n+1}$ be an $(n+1)\times(n+1)$ matrix all of whose variable elements are in $\mathfrak{X}$. Furthermore, let $S_{n}$ be the $n\times n$ 
submatrix of $M_{n+1}$ which is obtained by deleting row $i$ and column $j$. Then the problem of finding the probability that $\mathrm{det}M_{n+1}$ undergoes the least change upon perturbation of element $m_{ij}$ (of $M_{n+1}$) is equivalent to the problem of finding the probability that $\det$\,$S_{n}=0$ if $S_{n}$ is of type $\mathbf{A}_n$ or $\mathbf{B}_n$, and to the problem of finding the probability that $\det$\,$S_{n}=1$ if $S_{n}$ is of type $\mathbf{C}_n$.
\end{prop}
{\bf Proof}\ \ \ Expand the determinant of $S_{n}$ via the permutation group $\mathfrak{S}_n$ on the set $\{1,2,3,\ldots,n\}$:
\begin{equation}\nonumber
\displaystyle \mathrm{det}\, S_{n}=\sum_{\sigma\in\mathfrak{S}_{n}}sgn(\sigma)s_{\sigma(1)1}s_{\sigma(2)2}\cdots s_{\sigma(n)n},
\end{equation}
where $sgn(\sigma)$ is the sign of permutation $\sigma$ ($sgn(\sigma)=1$ if $\sigma$ is an even permutation, and $sgn(\sigma)=-1$ if $\sigma$ is odd). 
First, consider a matrix $S_{n}$ of type $\mathbf{A}_n$ or $\mathbf{B}_n$.
\textsl{Claim~1}   Let $x_{0}$ be a non-zero number. If $S_{n}$ is of type $\mathbf{A}_n$ or $\mathbf{B}_n$, then the probability that $\det S_{n}=x_{0}$ is 0.
\textsl{Proof of Claim~1}   If $\det S_{n}=x_{0}$, then at least one term in the expansion of $\det S_{n}$ is non-zero, and every element of $S_{n}$ that occurs in at least one non-zero term in that expansion is non-zero. Since $S_{n}$ is of type $\textbf{A}_n$ or $\textbf{B}_n$, every non-zero term in the expansion of $\det S_{n}$ contains at least one variable element of $S_{n}$. Because $\det S_{n}=x_{0}$, the sum of all the non-zero terms must be $x_{0}$, so the value $y_{0}$ of the ``last'' non-zero variable element is uniquely determined from all the others. For example, if $n=2$ and $S_{2}$ is of type $\mathbf{A}_2$, then all the elements of $S_{2}$ are variable. Because $S_{2}$ is a $2\times 2$ matrix with a non-zero determinant and has at least one non-zero term in the expansion of its determinant, at least one of the products $s_{11}s_{22},\ s_{21}s_{12}$ is non-zero. Suppose that $s_{11}s_{22}$ is non-zero. If we consider $s_{22}$ to be the last element, then\[\det S_{2}=x_{0}\implies(s_{11})(s_{22})-(s_{21})(s_{12})=x_{0}\implies s_{22}=y_{0}=\frac{x_{0}+(s_{21})(s_{12})}{s_{11}}\]
If $y_{0}\notin\mathfrak{X}$, it is obvious that $P(X=y_{0})=0$, so assume that $y_{0}\in\mathfrak{X}$. Since $\mathfrak{X}$ is a continuous set, it follows from (\ref{FeatureProbOnContinuousX_p2}) that $P(X=y_{0}$)=0.  From this result, together with the usual rules for computing the probability that a finite sum of finite products of numbers (such as the expansion of $\det S_{n}$ via $\mathfrak{S}_n$) has a certain value, it follows that the probability that $\det S_{n}=x_{0}$ is 0.  \textsl{End of Proof of Claim~1}
By Claim~1, the probability that $\det M_{n+1}$ will undergo the least change upon perturbation of element $m_{ij}$ is equal to the probability that $\det S_{n}=0$.
\textsl{Claim~2}   If $S_{n}$ is of type $\mathbf{A}_n$ or $\mathbf{B}_n$, then the probability that $\det S_{n}=0$ and at least one term in the expansion of $\det S_{n}$ via $\mathfrak{S}_n$ is non-zero is 0.
\textsl{Proof of Claim~2}   If at least one term in the expansion of $\det S_{n}$ is non-zero, then every element of $S_{n}$ that occurs in at least one non-zero term in that expansion is non-zero. Since $S_{n}$ is of type $\textbf{A}_n$ or $\textbf{B}_n$, every non-zero term in the expansion of $\det S_{n}$ contains at least one variable element of $S_{n}$. Because $\det S_{n}=0$, the sum of all the non-zero terms must be $0$, so the value $y_{0}$ of the ``last'' non-zero variable element is uniquely determined from all the others. Just as in the proof of Claim~1, the probability that that last non-zero variable element has a value of $y_{0}$ is 0, and the probability that $\det S_{n}=0$ and at least one term in the expansion of $\det S_{n}$ via $\mathfrak{S}_n$ is non-zero is 0.  \textsl{End of Proof of Claim~2}.
From Claims~1 and~2, it follows that if $S_{n}$ is of type $\mathbf{A}_n$ or $\mathbf{B}_n$, then the probability that $\det M_{n+1}$ undergoes the least change upon perturbation of element $m_{ij}$ is equal to the probability that {\it every} term in the expansion of $\det S_{n}$ is 0.
Now consider a matrix $S_{n}$ of type $\mathbf{C}_n$. We will refer to the term $s_{11}\cdots s_{nn}$ in the expansion of $\det S_{n}$ via $\mathfrak{S}_n$ as the {\it diagonal} term, and to each of the other terms as a {\it non-diagonal} term. Note that the diagonal term cannot be 0, since all the diagonal elements of $S_{n}$ are fixed at 1. If $x_{0}$ is a number other than $1$, then by reasoning analogous to that which was used in the proof of Claim~1, the probability that $\det S_{n}=x_{0}$ is 0. Also, by reasoning similar to that which was used in the proof of Claim~2, the probability that $\det S_{n}=1$ and at least one non-diagonal term in the expansion of $\det S_{n}$ via $\mathfrak{S}_n$ is non-zero is 0. Thus if $S_{n}$ is of type $\textbf{C}_n$, the probability that $\det M_{n+1}$ undergoes the least change upon perturbation of element $m_{ij}$ is equal to the probability that {\it every} non-diagonal term in the expansion of $\det S_{n}$ is 0.$\blacksquare$
We will use the term {\it binary matrix} to mean a matrix of 0's and 1's.
For a matrix $S_{n}$ of type $\mathbf{A}_n,\ \mathbf{B}_n$, or $\mathbf{C}_n$, let $\tilde{S}_{n}=\{\tilde{s}_{ij}\}$ be the $n\times n$ binary matrix with $\tilde{s}_{ij}=1$ if $s_{ij}\ne 0$, and $\tilde{s}_{ij}=0$ if $s_{ij}=0$. We will say that $\tilde{s}_{ij}$ is a {\it variable} (resp.\ {\it fixed}) element of $\tilde{S}_{n}$ if $s_{ij}$ is a variable (resp.\ fixed) element of $S_{n}$, and that $\tilde{S}_{n}$ is of type $\tilde{\mathbf{A}}_{n}$ (resp.\ $\tilde{\mathbf{B}}_{n},\ \tilde{\mathbf{C}}_{n}$) if $S_{n}$ is of type $\mathbf{A}_n$ (resp.\ $\mathbf{B}_n,\ \mathbf{C}_n$).  Every variable element of $\tilde{S}_n$ has probability $r$ of being 1, and probability $1-r$ of being 0.
Denote the permanent of any matrix $A$ by $\mathrm{per}\, A$. The expansion of the permanent of $\tilde{S}_{n}$ via $\mathfrak{S}_n$ is
\begin{equation}\label{DefPermanent_p2}
\displaystyle \mathrm{per}\,\tilde{S}_{n}=\sum_{\sigma\in\mathfrak{S}_{n}}\tilde{s}_{\sigma(1)1}\tilde{s}_{\sigma(2)2}\cdots\tilde{s}_{\sigma(n)n}
\end{equation}
Since $\tilde{S}_{n}$ is a binary matrix, every term in this expansion is either 0 or 1, so the value of $\mathrm{per}\,\tilde{S}_{n}$ is a non-negative integer.
The next result follows from Proposition~\ref{prop1_p2}   together with techniques analogous to those used in the proof of it.
\begin{prop}\label{prop2_p2}
Let $u$ be a real number. If $M_{n+1},\ S_{n}$, and $\mathfrak{X}$ are as hypothesized in Proposition~\ref{prop1_p2}, then the probability that $\det S_{n}=u$ is equal to the probability that $\mathrm{per}\,\tilde{S}_{n}=u$. Thus finding the probability that $M_{n+1}$ undergoes the least change upon perturbation of element $m_{ij}$ (of $M_{n+1})$ is equivalent to the problem of finding the probability that $\mathrm{per}\,\tilde{S_{n}}=0$ if $\tilde{S_{n}}$ is of type $\tilde{\mathbf{A}}_{n}$ or $\tilde{\mathbf{B}}_{n}$, and to the problem of finding the probability that $\mathrm{per}\,\tilde{S_{n}}=1$ if $\tilde{S_{n}}$ is of type $\tilde{\mathbf{C}}_{n}$.
\end{prop}
Throughout section~\ref{SecSolutions_p2}, we assume $\mathfrak{X}$ to be a continuous set such that $0\in\mathfrak{X}$.  
More will be said about this assumption in section~\ref{SecCointinuousXandDiscreteX_p2}.  
\section{Continuous $\mathfrak{X}$}\label{SecSolutions_p2}
Let $\tilde{S}_{n}$ be a matrix of type 
$\tilde{\mathbf{A}}_{n},\ \tilde{\mathbf{B}}_{n}$, or $\tilde{\mathbf{C}}_{n}$, so that some of its elements are variable, its other elements are fixed at 1, and each of its variable elements has probability $r$ of being 1 and probability $1-r$ of being 0. In this section, we formulate the probability that $\mathrm{per}$\,$\tilde{S}_{n}=u$, where $u=0$ if $\tilde{S_{n}}$ is of type $\tilde{\mathbf{A}}_{n}$ or $\tilde{\mathbf{B}}_{n}$, and $u=1$ if $\tilde{S_{n}}$ is of type $\tilde{\mathbf{C}}_{n}$.   
In what follows, we will use the expression $\mathrm{per}\,\tilde{S}_{n}=u$ to mean that $\mathrm{per}\,\tilde{S}_{n}=0$ if $\tilde{S}_{n}$ is of type $\tilde{\mathbf{A}}_{n}$ or $\tilde{\mathbf{B}}_{n}$, and that $\mathrm{per}\,\tilde{S}_{n}=1$ if $\tilde{S}_{n}$ is of type $\tilde{\mathbf{C}}_{n}$. Also, we will say that a matrix $\tilde{S}_{n}$ of a given type ($\tilde{\mathbf{A}}_{n},\ \tilde{\mathbf{B}}_{n}$, or $\tilde{\mathbf{C}}_{n}$) is a {\it pertinent} matrix of that type if $\mathrm{per}\,\tilde{S}_{n}=u$.
For a matrix $\tilde{S}_{n}$ of a given type, let $m$ be the number of variable elements in $\tilde{S}_{n}$, and let $i_{\max}$ be the maximum number of variable elements with a value of 1 that a pertinent matrix of that type can have.  Then the probability that $\mathrm{per}$\,$\tilde{S}_{n}=u$ is 
\begin{equation}\label{PS_p2}
P_{\mathrm{per}\,\tilde{S}_{n}=u}(r)=\displaystyle \sum_{i=0}^{i_{\max}}E_{n}(i)\cdot r^{i}\cdot(1-r)^{m-i},
\end{equation}
where $E_{n}(i)$ is the number of pertinent matrices of that type which have exactly $i$ variable elements with a value of 1---except for type $\tilde{\textbf{B}}_{n}$, where $E_{n}(i)$ is the number of pertinent matrices which have exactly $i$ variable elements with a value of 1 and  whose special diagonal element $b_{11}$ is in the same location as the special diagonal element of $\tilde{S}_{n}$.  
The values of $m$ and $i_{\max}$ depend on the type of matrix. Clearly, the value of $m$ for type $\tilde{\mathbf{A}}_{n}$ (resp.\ $\tilde{\mathbf{B}}_{n},\ \tilde{\mathbf{C}}_{n}$) is $n^{2}$ (resp.\ $n^{2}-n+1,$\ $n^{2}-n$).  
To determine the value of $i_{\max}$ for a given type, it may be easier to first find the minimum number $j_{\min}$ of variable elements with a value of 0 that a pertinent matrix of that type can have, and then use the relation $i_{\max}=m-j_{\min}$ to compute $i_{\max}$.
There exist pertinent matrices $\tilde{S}_{n}$ of types $\tilde{\mathbf{A}}_{n}$ and $\tilde{\mathbf{B}}_{n}$ in which all $n$ variable elements in a single row or column have a value of 0 and all the other variable elements have a value of 1. It can easily be shown by induction on $n$ that $\mathrm{per}\,\tilde{S}_{n}>0$ if $\tilde{S}_{n}$ is of type $\tilde{\mathbf{A}}_{n}$ or $\tilde{\mathbf{B}}_{n}$ and has no row or column comprised of $n$ variable elements with a value of 0. Thus $j_{\min}=n$ (and $i_{\max}=m-n$) for types $\tilde{\mathbf{A}}_{n}$ and $\tilde{\mathbf{B}}_{n}$.  
In a matrix of type $\tilde{\mathbf{A}}_{n}$, any of the $n$ rows or $n$ columns could be the single row or column comprised entirely of variable elements with a value of 0, so the total number of candidates for that row or column is $n+n=2n$ (except for $n=1$, where that row and column coincide). Thus $E_{n}(i_{\max})=2n$ if $n>1$, and $E_{1}(i_{\max})=1$.
In a matrix of type $\tilde{\mathbf{B}}_{n}$ with special diagonal element $b_{11}$, there are just two candidates for the single row or column comprised entirely of variable elements with a value of 0 (namely, row $1$ and column $1$), the only exception being $n=1$, where row $1$ and column $1$ coincide. Thus $E_{n}(i_{\max})=2$ if $n>1$, and $E_{1}(i_{\max})=1$.
If a matrix of type $\tilde{\mathbf{C}}_{n}$ is in upper-triangular form (meaning that every element below the main diagonal is 0) and all the elements above the main diagonal have a value of 1, then $\mathrm{per}\,\tilde{S}_{n}=1$. Similarly, if a matrix of type $\tilde{\mathbf{C}}_{n}$ is in lower-triangular form (meaning that every element above the main diagonal is 0) and all the elements below the main diagonal have a value of 1, then $\mathrm{per}\,\tilde{S}_{n}=1$.  Such a matrix has $(n^{2}-n)/2$ variable elements with a value of 0, so $j_{\min}\le(n^{2}-n)/2$. 
In section~\ref{SecCase(iii)_p2}  we show that $j_{\min}=(n^{2}-n)/2$ for type $\tilde{\mathbf{C}}_{n}$.  However, we will see that $E_{n}(i_{\max})>2$ if $n>2$, so there are pertinent matrices of type $\tilde{\mathbf{C}}_{n}$ which are in neither upper- nor lower-triangular form but have $i_{\max}$ variable elements with a value of 1.
What remains to be done is to evaluate $E_{n}(i)$ for each type of matrix, which entails counting matrices with $\mathrm{per}\,\tilde{S}_{n}=u$.
\subsection{Case (i): Type $\tilde{\mathbf{A}}_{n}$}\label{SecCase(i)_p2}
Denote $\tilde{S}_{n}$ by $\tilde{A}_{n}$, and $E_{n}(i)$ by $F_{n}(i)$.  Since $m=n^{2}$ and $j_{\min}=n$, we see that $i_{\max}=n^{2}-n$, so the probability that $\mathrm{per}\,\tilde{A}_{n}=0$ is
\begin{equation}
\label{PA_p2}
P_{\mathrm{per}\,\tilde{A}_{n}=0}(r)=\displaystyle \sum_{i=0}^{n^{2}-n}F_{n}(i)\cdot r^{i}\cdot(1-r)^{n^{2}-i}
\end{equation}
Thus to evaluate $P_{\mathrm{per}\,\tilde{A}_{n}=0}(r)$, we have to find the integer sequence $\{F_{n}(i)\}_{i=0}^{i=n^{2}-n}$.
The values of $F_{n}(i)$ for $n=1,\ldots,4$ (and $i=0,\ldots,n^{2}-n$) are given in Table~\ref{FnRuleTable_p2}.  For $n=5$, the sequence (from $i=0$ through $i=n^{2}-n=5^{2}-5=20$) is 1, 25, 300, 2300, 12650, 53010, 174700, 458500, 956775, 1571525, 2010920, 1994200, 1534800, 923700, 439600, 166720, 50025, 11500, 1900, 200, 10.
\begin{table}[htbp]
\scriptsize
\begin{center}
\caption{Values of $F_{n}(i)$ for $n=1,\ldots,4$ (and $i=0,\ldots,n^{2}-n$)}
\begin{tabular}{|c|c|c|c|c|c|c|c|c|c|c|c|c|c|}\hline
$n\backslash i$&0&1&2&3&4&5&6&7&8&9&10&11&12\\ \hline
    1&1&&&&&&&&&&&&\\ \hline
    2&1&4&4&&&&&&&&&&\\ \hline
    3&1&9&36&78&90&45&6&&&&&&\\ \hline
    4&1&16&120&560&1796&4080&6496&6976&4860&2128&576&96&8\\ \hline
\end{tabular}
\label{FnRuleTable_p2}
\end{center}
\end{table}
See Sloane's sequence A088672~\cite{Sloane}  for the first few terms of the sequence $\displaystyle \left\{f_{n}\right\}=\left\{\sum_{i}F_{n}(i)\right\}$ (i.e., the first few terms of the sequence 1, 9, 265, 27713, 10363661,$\ldots$).  Its asymptotic behavior has been proposed by Everett and Stein~\cite{EverettStein}    to be $2^{(n^{2}-n+1)}$.
\subsection{Case (ii): Type $\tilde{\mathbf{B}}_{n}$}\label{SecCase(ii)_p2}
Denote $\tilde{S}_{n}$ by $\tilde{B}_{n}$, and $E_{n}(i)$ by $G_{n}(i)$.  Since $m=n^{2}-n+1$ and $j_{\min}=n$, we see that $i_{\max}=(n^{2}-n+1)-n=(n-1)^{2}$, so the probability that $\mathrm{per}\,\tilde{B}_{n}=0$ is
\begin{equation}\label{PB_p2}
P_{\mathrm{per}\,\tilde{B}_{n}=0}(r)=\displaystyle \sum_{i=0}^{(n-1)^{2}}G_{n}(i)r^{i}(1-r)^{n^{2}-n+1-i}
\end{equation}
Thus to evaluate $P_{\mathrm{per}\,\tilde{B}_{n}=0}(r)$, we have to find the integer sequence $\{G_{n}(i)\}_{i=0}^{i=(n-1)^{2}}$.
The values of $G_{n}(i)$ for $n=1,\ldots,4$ (and $i=0,\ldots,(n-1)^{2}$) are given in Table~\ref{GnRuleTable_p2}.  For $n=5$, the sequence (from $i=0$ through $i=(n-1)^{2}=(5-1)^{2}=16$) is 1, 20, 186, 1056, 4035, 10836, 21032, 30212, 32829, 27520, 18062, 9324, 3741, 1128, 240, 32, 2.  At this time, the sequence $\displaystyle \{g_{n}\}=\left\{\sum_{i}G_{n}(i)\right\}$ is not given in Sloane's list~\cite{Sloane}.
\begin{table}[htbp]
\scriptsize
\begin{center}\caption{Values of $G_{n}(i)$ for $n=1,\ldots,4$ (and $i=0,\ldots,(n-1)^{2}$)}
\begin{tabular}{|c|c|c|c|c|c|c|c|c|c|c|c|}\hline
$n\backslash i$&0&1&2&3&4&5&6&7&8&9\\ \hline
    1&1&&&&&&&&&\\ \hline
    2&1&2&&&&&&&&\\ \hline
    3&1&6&13&10&2&&&&&\\ \hline
    4&1&12&63&184&315&324&203&78&18&2\\ \hline
\end{tabular}
\label{GnRuleTable_p2}
\end{center}
\end{table}
\subsection{Case (iii): Type $\tilde{\mathbf{C}}_{n}$}\label{SecCase(iii)_p2}
Denote $\tilde{S}_{n}$ by $\tilde{C}_{n}$, and $E_{n}(i)$ by $H_{n}(i)$.  The probability that $\mathrm{per}\,\tilde{C}_{n}=1$ is
\begin{equation}\label{PC_p2}
P_{\mathrm{per}\,\tilde{C}_{n}=1}(r)=\displaystyle \sum_{i=0}^{i_{\max}}H_{n}(i)r^{i}(1-r)^{n^{2}-n-i}
\end{equation}
Thus to evaluate $P_{\mathrm{per}\,\tilde{C}_{n}=1}(r)$, we have to determine the value of $i_{\max}$ and find the integer sequence $\{H_{n}(i)\}_{i=0}^{i=i_{\max}}$.
\begin{prop}\label{prop3_p2}
The number of matrices of type $\tilde{\mathbf{C}}_{n}$ with permanent 1 and $i$ variable elements with a value of 1 is equal to the number of acyclic digraphs with vertex set $\{1,2,3,\ldots,n\}$ and $i$ edges.  (The term {\it digraph} is short for directed graph; an acyclic digraph is a digraph that contains no cycles.)
\end{prop}
{\bf Proof}\ \ \ For every matrix $\tilde{S}_{n}$ of type $\textbf{C}_n$, let $G(\tilde{S}_{n})$ be the digraph with vertex set $\{1,2,3,\ldots,n\}$ and adjacency matrix $\tilde{S_{n}}-I_{n}$, where $I_{n}$ is the $n\times n$ identity matrix. Then the edge set of $G(\tilde{S}_{n})$ is $\{(k,l):k\ne l,\ \tilde{s}_{kl}=1\}$, and the number of edges of $G(\tilde{S}_{n})$ is equal to the number of variable elements of $\tilde{S}_{n}$ with a value of 1.
\textsl{Claim}   $\mathrm{per}\,\tilde{S}_{n}=1$ if and only if $G(\tilde{S}_{n})$ is an acyclic digraph.
\textsl{Proof of Claim}   All the diagonal elements of $\tilde{S}_{n}-I_{n}$ are 0, so $G(\tilde{S}_{n})$ has no loops (a loop is an edge that connects a vertex to itself). If $G(\tilde{S}_{n})$ has a cycle, then there exist $j$ (with $2\le j\le n$) and distinct vertices $k_{1},\ldots,k_{j}$ of~$G(\tilde{S}_{n})$ such that $(k_{1},k_{2}),(k_{2},k_{3}),\ldots,(k_{j-1},k_{j}),(k_{j},k_{1})$ are edges of $G(\tilde{S}_{n})$. Thus $\tilde{s}_{k_{1}k_{2}},\tilde{s}_{k_{2}k_{3}},\ldots,\tilde{s}_{k_{j-1}k_{j}},\tilde{s}_{k_{j}k_{1}}$ are variable elements of $\tilde{S}_{n}$ that have a value of~1. If $j=n$, then $\tilde{s}_{k_{1}k_{2}}\cdot\tilde{s}_{k_{2}k_{3}}\cdots\tilde{s}_{k_{n-1}k_{n}}\cdot\tilde{s}_{k_{n}k_{1}}$ is a non-diagonal term in the expansion of $\mathrm{per}\tilde{S}_{n}$ via $\mathfrak{S}_n$ which is equal to 1. If $j<n$, let $l_{1},l_{2},\ldots,l_{n-j}$ be the elements of $\{1,2,3,\ldots,n\}$ which are not in the set $\{k_{1},\ldots,k_{j}\}$. Then $\tilde{s}_{k_{1}k_{2}}\cdot\tilde{s}_{k_{2}k_{3}}\cdots\tilde{s}_{k_{j-1}k_{j}}\cdot\tilde{s}_{k_{j}k_{1}}\cdot\tilde{s}_{l_{1}l_{1}}\cdot\tilde{s}_{l_{2}l_{2}}\cdots\tilde{s}_{l_{n-j}l_{n-j}}$ is a non-diagonal term in the expansion of $\mathrm{per}\tilde{S}_{n}$ via $\mathfrak{S}_n$ which is equal to 1. In either case, $\mathrm{per}\tilde{S}_{n}>1$.
If $G(\tilde{S}_{n})$ is acyclic, then the only non-zero term in the expansion of $\mathrm{per}\tilde{S}_{n}$ via $\mathfrak{S}_n$ is the diagonal term, so $\mathrm{per}\tilde{S}_{n}=1$.  \textsl{End of Proof of Claim}
Clearly, $G$ is a one-to-one function. Thus by the Claim and the definition of $G$, the number of matrices of type $\tilde{\mathbf{C}}_{n}$ with permanent 1 and $i$ variable elements with a value of 1 is equal to the number of acyclic digraphs with vertex set $\{1,2,3,\ldots,n\}$ and $i$ edges. $\blacksquare$
Now suppose that a matrix $\tilde{S}_{n}$ of type $\tilde{\mathbf{C}}_{n}$ has fewer than $(n^{2}-n)/2$ variable elements with a value of 0. Since $\tilde{S}_{n}$ has a total of $n^{2}-n$ variable elements, it must have more than $(n^{2}-n)/2$ elements with a value of 1. Thus there are $k,l$ such that $1\le k<l\le n$, and (variable) elements $\tilde{s}_{kl}$ and $\tilde{s}_{lk}$ of $\tilde{S}_{n}$ that have a value of 1. But then the graph $G(\tilde{S}_{n})$ has a cycle of length 2, namely, $(k,l),\,(l,k)$. By the Claim, $\mathrm{per}\tilde{S}_{n}>1$. From this it follows that a pertinent matrix of type $\tilde{\mathbf{C}}_{n}$ cannot have fewer than $(n^{2}-n)/2$ variable elements with a value of 0. As shown earlier, there exist pertinent matrices of type $\tilde{\mathbf{C}}_{n}$ with exactly $(n^{2}-n)/2$ variable elements that have a value of 0. Thus $j_{\min}=(n^{2}-n)/2$ (and $i_{\max}=(n^{2}-n)-[(n^{2}-n)/2]=(n^{2}-n)/2$).
By Proposition~\ref{prop3_p2}, the problem of evaluating $H_{n}(i)$, the number of matrices of type $\tilde{\mathbf{C}}_{n}$ with permanent 1 and $i$ variable elements with a value of 1, can be transformed into the problem of counting acyclic digraphs with vertex set $\{1,2,3,\ldots,n\}$ and $i$ edges. We can use a result from graph theory to find $H_{n}(i)$.  I. M. Gessel~\cite{Gessel}     counted labeled acyclic digraphs according to the number of sources, sinks, and edges.  (A source is a vertex with in-degree 0, and a sink is a vertex with out-degree 0.)  We use that result in the proof of Proposition~\ref{prop4_p2}    below.
\begin{prop}\label{prop4_p2}
\begin{eqnarray}\label{HnForm1_p2}
H_{n}(e)&=&\displaystyle \frac{1}{e!}\cdot\left.\frac{d^{e}Y_{n}(t)}{dt^{e}}\right|_{t=0}\\
\label{HnForm2_p2}
Y_{n}(t)&=&(1+t)^{{}_{n}C_{2}}\displaystyle \left.\left[\frac{\partial^{n}}{\partial z^{n}}\left(\frac{1}{Z(-z,t)}\right)\right]\right|_{z=0}\\
\label{HnForm4_p2}
Z(z,t)&=&\displaystyle \sum_{n=0}^{\infty}\frac{z^{n}}{n!\cdot(1+t)^{{}_{n}C_{2}}}
\end{eqnarray}
\end{prop}
{\bf Proof}\ \ \ For an acyclic digraph $D_{n}$ with vertex set $\{1,2,\ldots,n\}$, let $e(D_{n})$ and $s(D_{n})$ be the numbers of edges and sources, respectively, of $D_{n}$.  Then define a function $Y_{n}(t;\alpha)$ as
\begin{equation}\nonumber
Y_{n}(t;\displaystyle \alpha)=\sum_{D_{n}}t^{e(D_{n})}\cdot\alpha^{s(D_{n})},
\end{equation}
where the sum is over all acyclic digraphs $D_{n}$. (Throughout this proof, an expression of the form $a^{b}$ is to be interpreted as 1 if both $a$ and $b$ are 0. For example, $t^{0}$ is to be interpreted as 1 if $t=0$, and $\alpha^{0}$ is to be interpreted as 1 if $\alpha=0$.) 
Let $p_{n}(e,s)$ be the number of $D_{n}$ with $e$ edges and $s$ sources.  Then
\begin{equation}\nonumber
p_{n}(e,s)=\displaystyle \frac{1}{e!s!}\cdot\left[\frac{\partial^{e}}{\partial t^{e}}\left.\frac{\partial^{s}}{\partial\alpha^{s}}Y_{n}(t;\alpha)\right]\right|_{t=\alpha=0}
\end{equation}
and we can express $Y_{n}(t;\alpha)$ as
\begin{equation}\nonumber
Y_{n}(t;\displaystyle \alpha)=\sum_{e}\sum_{s}p_{n}(e,s)\cdot t^{e}\cdot\alpha^{s}
\end{equation}
Denoting $Y_{n}(t;1)$ by $Y_{n}(t)$, we have
\begin{equation}\nonumber
Y_{n}(t)=\displaystyle \sum_{e}\left(\sum_{s}p_{n}(e,s)\right)\cdot t^{e}=\sum_{e}H_{n}(e)\cdot t^{e},
\end{equation}
where $H_{n}(e)$ is the number of $D_{n}$ with $e$ edges. Thus
\begin{equation}\nonumber
H_{n}(e)=\displaystyle \left.\frac{1}{e!}\cdot\frac{d^{e}Y_{n}(t)}{dt^{e}}\right|_{t=0}
\end{equation}
What remains is to find a formula for $Y_{n}(t)$ that will enable us to evaluate $H_{n}(e)$.
The following relation is proved in~\cite{Gessel}:
\begin{equation}\label{HnProof1_p2}
\displaystyle \sum_{j=0}^{n}(1+t)^{j(n-j)}{}_{n}C_{j}\alpha^{j}Y_{n-j}(t;1)=Y_{n}(t;\alpha+1)
\end{equation}
It is well known that
\begin{eqnarray}
&&\displaystyle \sum_{j=0}^{n}(1+t)^{j(n-j)}\,{}_{n}C_{j}\, a_{j}b_{n-j}=c_{n}\\\nonumber
&\Leftrightarrow&\left[\displaystyle \sum_{n=0}^{\infty}a_{n}\frac{z^{n}}{n!\cdot(1+t)^{{}_{n}C_{2}}}\right]\left[\sum_{n=0}^{\infty}b_{n}\frac{z^{n}}{n!\cdot(1+t)^{{}_{n}C_{2}}}\right]=\left[\sum_{n=0}^{\infty}c_{n}\frac{z^{n}}{n!\cdot(1+t)^{{}_{n}C_{2}}}\right]
\end{eqnarray}
Setting $a_{n},\ b_{n}$, and $c_{n}$ to $\alpha^{n},\ Y_{n}(t;1)$ ($=Y_{n}(t)$), and $Y_{n}(t;\alpha+1)$, respectively, we see that (\ref{HnProof1_p2}) is equivalent to
\begin{equation}\label{XYrelation1_p2}
\displaystyle \sum_{n=0}^{\infty}\frac{(\alpha z)^{n}}{n!\cdot(1+t)^{{}_{n}C_{2}}}\cdot\sum_{n=0}^{\infty}\frac{Y_{n}(t)\cdot z^{n}}{n!\cdot(1+t)^{{}_{n}C_{2}}}=\sum_{n=0}^{\infty}\frac{Y_{n}(t;\alpha+1)\cdot z^{n}}{n!\cdot(1+t)^{{}_{n}C_{2}}}
\end{equation}
There is only one acyclic digraph with 0 vertices (namely, the empty graph, which has no edges and no sources), and for $n\ge 1$ every acyclic digraph with $n$ vertices has at least one source. Thus 
\begin{eqnarray*}
&&Y_{0}(t;0)=p_{0}(0,0)\cdot t^{0}\cdot 0^{0}=1\\
n\displaystyle \ge 1&\Rightarrow&Y_{n}(t;0)=\displaystyle \sum_{e}\left(\sum_{s\ge 1}p_{n}(e,s)\cdot 0^{s}\right)\cdot t^{e}=0
\end{eqnarray*}
Setting $\alpha$ to $-1$ in (\ref{XYrelation1_p2}) and defining $Z(z,t)$ as $\displaystyle \sum_{n=0}^{\infty}\frac{z^{n}}{n!\cdot(1+t)^{{}_{n}C_{2}}}$, we find that
\begin{equation}\nonumber
\displaystyle \sum_{n=0}^{\infty}Y_{n}(t)\cdot\frac{z^{n}}{n!\cdot(1+t)^{{}_{n}C_{2}}}=\frac{1}{Z(-z,t)},
\end{equation}
Thus we obtain (\ref{HnForm2_p2}).  $\blacksquare$
{\bf Example}\ \ \ We show the calculation of $H_{4}(i)$ as an example.  By (\ref{HnForm4_p2}),
\begin{equation}\nonumber
 Z(-z,t)=1-z+\displaystyle \frac{z^{2}}{2(1+t)}-\frac{z^{3}}{6(1+t)^{3}}+\frac{z^{4}}{24(1+t)^{6}}-\frac{z^{5}}{120(1+t)^{10}}+\cdots
\end{equation}
In the calculation of $Y_{4}(t)$ (see (\ref{HnForm2_p2})), we take the fourth partial derivative of $1/Z(-z,t)$ with respect to $z$ and evaluate it at $z=0$:
\begin{equation}\nonumber
\displaystyle \left.\left[\frac{\partial^{4}}{\partial z^{4}}\left(\frac{1}{1-x+\frac{z^{2}}{2(1+t)}-\frac{z^{3}}{6(1+t)^{3}}+\frac{z^{4}}{24(1+t)^{6}}-\frac{z^{5}}{120(1+t)^{10}}+\cdots}\right)\right]\right|_{z=0}
\end{equation}
The terms of $Z(-z,t)$ that contain a factor of $z^{k}$ for some $k\ge 5$ will vanish when the fourth partial derivative of $1/Z(-z,t)$ is evaluated at $z=0$, so it suffices to use the following finite sum, $Z_{4}(-z,t)$, in place of $Z(-z,t)$ when we perform the differentiation:
\begin{equation}\nonumber
Z_{4}(-z,t)=1-z+\displaystyle \frac{z^{2}}{2(1+t)}-\frac{z^{3}}{6(1+t)^{3}}+\frac{z^{4}}{24(1+t)^{6}}
\end{equation}
Thus
\begin{eqnarray}\nonumber
 Y_{4}(t)&=&(1+t)^{{}_{4}\mathrm{C}_{2}}\displaystyle \cdot\left.\left[\frac{\partial^{4}}{\partial z^{4}}\left(\frac{1}{Z(-z,t)}\right)\right]\right|_{z=0}\\\nonumber
&\mbox{$=$}&\mbox{$\displaystyle (1+t)^{6}\cdot\left.\left[\frac{\partial^{4}}{\partial z^{4}}\left(\frac{1}{1-z+\displaystyle \frac{z^{2}}{2(1+t)}-\frac{z^{3}}{6(1+t)^{3}}+\frac{z^{4}}{24(1+t)^{6}}}\right)\right]\right|_{z=0}$}\\\nonumber
&=&1+12t+60t^{2}+152t^{3}+186t^{4}+108t^{5}+24t^{6}\\\nonumber
\end{eqnarray}
That is, ($H_{4}(0)$,\ $H_{4}(1)$,\ $H_{4}(2)$,\ $H_{4}(3)$,\ $H_{4}(4)$,\ $H_{4}(5)$,\ $H_{4}(6)$) = (1, 12, 60, 152, 186, 108, 24).  Thus
\begin{eqnarray}
\nonumber
&&P_{\mathrm{per}\,\tilde{C}_{4}=1}(r)=\displaystyle \sum_{i=0}^{6}H_{4}(i)r^{i}(1-r)^{12-i}\\\nonumber
&=&1\cdot(1-r)^{12}+12\cdot r^{1}(1-r)^{11}+60\cdot r^{2}(1-r)^{10}+152\cdot r^{3}(1-r)^{9}\\\nonumber
&&+186\cdot r^{4}(1-r)^{8}+108\cdot r^{5}(1-r)^{7}+24\cdot r^{6}(1-r)^{6}
\end{eqnarray}
$\blacksquare$
The values of $H_{n}(i)$ for $n=1,\ldots,5$ (and $i=0,\ldots,(n^{2}-n)/2$) are given in Table~\ref{HnRuleTable_p2}.  See Sloane's sequence A003024~\cite{Sloane}    for the first few terms of the sequence $\displaystyle \{h_{n}\}=\{\sum_{i}H_{n}(i)\}$ (i.e., the first few terms of the sequence 1, 3, 25, 543, 29281,$\ldots$).
\begin{table}[htbp]
\scriptsize
\begin{center}\caption{Values of $H_{n}(i), n=1,\ldots,5$ (and $i=0,\ldots,(n^{2}-n)/2$)}
\begin{tabular}{|c|c|c|c|c|c|c|c|c|c|c|c|}\hline
$n\backslash i$&0&1&2&3&4&5&6&7&8&9&10\\ \hline
    1&1&&&&&&&&&&\\ \hline
    2&1&2&&&&&&&&&\\ \hline
    3&1&6&12&6&&&&&&&\\ \hline
    4&1&12&60&152&186&108&24&&&&\\ \hline
    5&1&20&180&940&3050&6180&7960&6540&3330&960&120\\ \hline
\end{tabular}
\label{HnRuleTable_p2}
\end{center}
\end{table}
As shown in the table, there are six matrices of type $\tilde{\mathbf{C}}_{3}$ with permanent 1 and $i_{\max}$ ($=3$) variable elements that have a value of 1. Of those six matrices, there are four that are in neither upper- nor lower-triangular form:
\begin{equation}\nonumber
\left(\begin{array}{ccc}1&0&1\\
1&1&1\\0&0&1\end{array}\right),\\
\left(\begin{array}{ccc}1&0&0\\1&1&1\\
1&0&1\end{array}\right),\\
\left(\begin{array}{ccc}1&1&1\\0&1&0\\
0&1&1\end{array}\right),\\
\left(\begin{array}{ccc}1&1&0\\0&1&0\\
1&1&1\end{array}\right)
\end{equation}
\subsection{Remarks}\label{SecRemarks_p2}
\begin{figure}[h]\begin{center}
\begin{center}
\includegraphics[width=90mm,height=48.8mm]{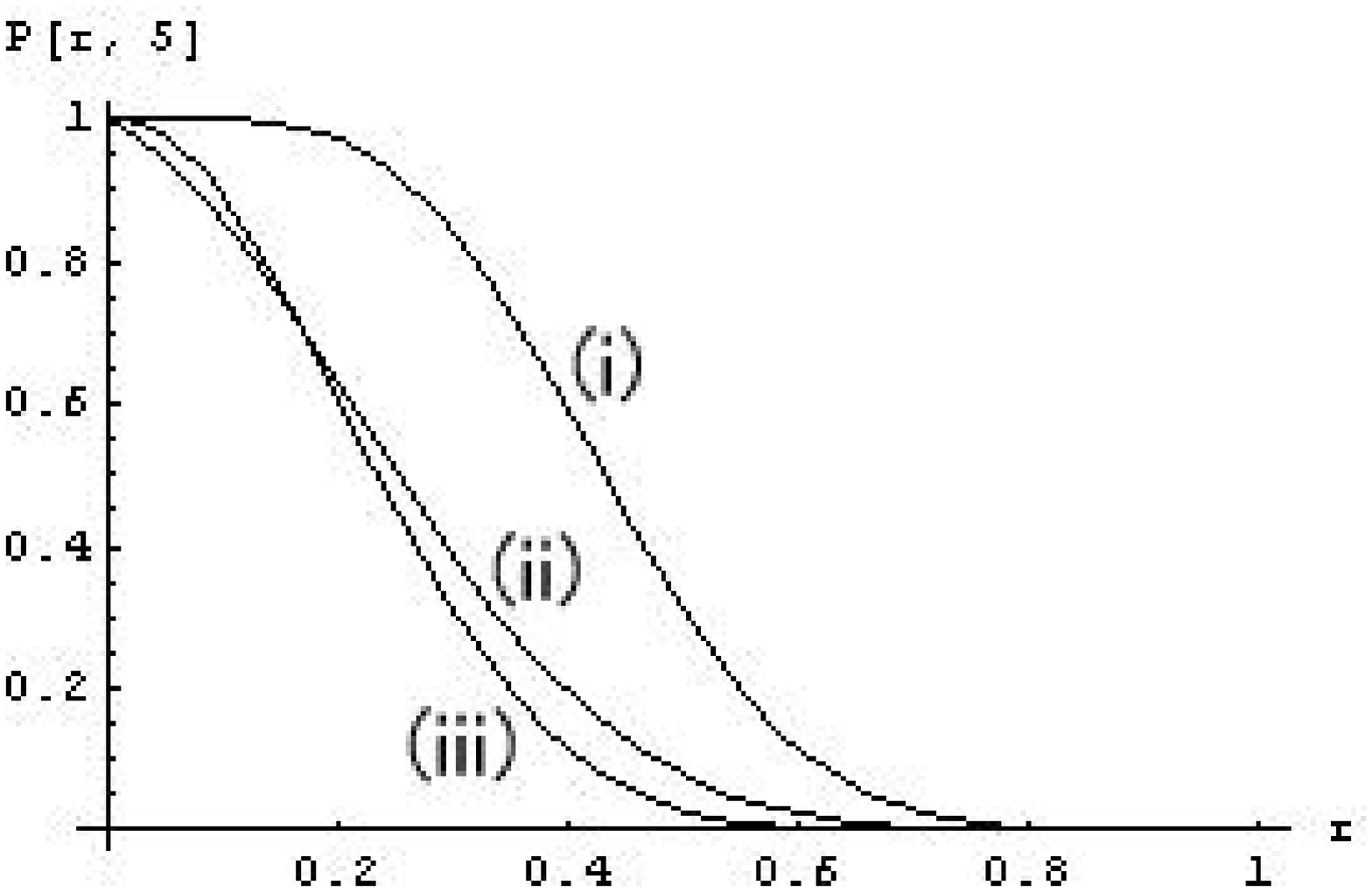}
\end{center}
\end{center}
\caption{(i) $P_{\mathrm{per}\,\tilde{A}_{5}=0}(r)$, (ii) $P_{\mathrm{per}\,\tilde{B}_{5}=0}(r)$, (iii) $P_{\mathrm{per}\,\tilde{C}_{5}=1}(r)$}\label{FigGraphs_p2}
\end{figure}

Graphs of the functions $P_{\mathrm{per}\,\tilde{A}_{5}=0}(r),\ P_{\mathrm{per}\,\tilde{B}_{5}=0}(r)$, and $P_{\mathrm{per}\,\tilde{C}_{5}=1}(r)$ vs.\ $r$ are presented in Fig.~\ref{FigGraphs_p2}.
If we were to start with a matrix of type $\tilde{\mathbf{A}}_{n}$ (case (i)) and replace all but one of the diagonal elements with a fixed element 1, we would obtain a matrix of type $\tilde{\textbf{B}}_{n}$ (case (ii)).  If we in turn replaced the special diagonal element in the latter matrix (the variable element $b_{11}$) with a fixed element 1, we would obtain a matrix of type $\tilde{\mathbf{C}}_{n}$ (case (iii)).  Therefore, we might expect to find that there exist analytic formulas for $P_{\mathrm{per}\,\tilde{B}_{n}=0}(r)$ in terms of $P_{\mathrm{per}\,\tilde{A}_{n}=0}(r)$, and $P_{\mathrm{per}\,\tilde{C}_{n}=0}(r)$ in terms of $P_{\mathrm{per}\,\tilde{B}_{n}=0}(r)$. At the very least, we might expect that the relation $P_{\mathrm{per}\,\tilde{A}_{n}=0}(r)>P_{\mathrm{p}\mathrm{e}\mathrm{r}\,\tilde{B}_{n}=0}(r)>P_{\mathrm{p}\mathrm{e}\mathrm{r}\,\tilde{C}_{n}=1}(r)$ would hold---and that the value of $P_{\mathrm{per}\,\tilde{A}_{n}=0}(r)-P_{\mathrm{per}\,\tilde{B}_{n}=0}(r)$ would be larger than the value of $P_{\mathrm{per}\,\tilde{B}_{n}=0}(r)-P_{\mathrm{per}\,\tilde{C}_{n}=1}(r)$. As can be seen in the figure, such  a simple relation holds for $n=5$, but only for values of $r$ larger than approximately 0.18. 
There is another relationship between matrices of type $\bm{\mathrm{B}}_{n}$ and those of type $\mathbf{C}_{n+1}$. 
It can be shown that if row $i$ and column $j$ are deleted from a matrix of type $\mathbf{C}_{n+1}$, the resulting $n\times n$ matrix can be converted to a matrix of type $\bm{\mathrm{B}}_{n}$ by interchange of at most one pair of rows or one pair of columns. It is well known that the absolute value of the determinant of a matrix is invariant under interchange of any two rows or any two columns. Thus the cofactor of any element of a matrix of type $\mathbf{C}_{n+1}$ is equal in absolute value to the determinant of a matrix of type $\bm{\mathrm{B}}_{n}$.  
Though we obtained a closed formula for the $n$th term of the sequence $\displaystyle \{h_{n}\}=\{\sum_{i}H_{n}(i)\}$ (where $h_{n}$ is the total number of matrices of type $\tilde{\mathbf{C}}_{n}$ with permanent 1), we did not obtain closed formulas for the $n$th term of the sequences $\displaystyle \{f_{n}\}=\{\sum_{i}F_{n}(i)\}$  and $\displaystyle \{g_{n}\}=\{\sum_{i}G_{n}(i)\}$ (where $f_{n}$ is the total number of matrices of type $\tilde{\mathbf{A}}_{n}$ with permanent 0, and $g_{n}$ is the number of matrices of type $\tilde{\mathbf{B}}_{n}$ with permanent 0 and a fixed location of the special diagonal element $b_{11}$).
We found that it took considerably more time to compute $\{f_{n}\}$ and $\{g_{n}\}$ than to compute $\{h_{n}\}$. It is possible that the problems of computing $\{f_{n}\},\ \{g_{n}\}$, and $\{h_{n}\}$ are related to some $\# P$-{\it complete} problem.  $\# P$-complete is a class of counting problems in complexity theory. The problem of computing the permanent of a binary matrix is $\# P$-{\it complete}, as reported by Valiant~\cite{Valiant}   and by Ben-Dor and Halevi~\cite{BenAmirHalevi}. The determinant of a given matrix can be computed in cubic time by Gaussian elimination, but that approach cannot be used to compute the permanent.  The key difference between determinants and permanents is that the relation $\mathrm{det}AB=\mathrm{d}\mathrm{e}\mathrm{t}A\cdot \mathrm{d}\mathrm{e}\mathrm{t}B$ always holds but the relation $\mathrm{per}AB=\mathrm{p}\mathrm{e}\mathrm{r}A\cdot \mathrm{p}\mathrm{e}\mathrm{r}B$ does not hold in general~\cite{Marcus}.
\section{Continuous $\mathfrak{X}$ vs.\ discrete $\mathfrak{X}$}\label{SecCointinuousXandDiscreteX_p2}
Let $u(\mathbf{A}_n,\mathfrak{X})$ be the real number $u$ of least absolute value such that a matrix of type $\mathbf{A}_n$ can have determinant $u$ with non-zero probability if its variable elements are chosen from $\mathfrak{X}$. 
If such $u$ exists and $\mathfrak{X}$ is discrete, let $\Omega(\mathbf{A}_n,\mathfrak{X})$ be the set of all matrices $S_{n}$ of type $\mathbf{A}_n$ such that $\det S_{n}=u(\mathbf{A}_n,\mathfrak{X})$ with non-zero probability.
If such $u$ exists and $\mathfrak{X}$ is continuous, define the following equivalence relation on the set of matrices $A_{n}$ of type $\mathbf{A}_n$ such that $\det A_{n}=u(\mathbf{A}_n,\mathfrak{X})$ with non-zero probability: Matrices $M$ and $N$ are equivalent if the locations of the variable elements that have a non-zero value are identical in $M$ and $N$. Then let $\Omega(\mathbf{A}_n,\mathfrak{X})$ be a set that contains one representative of each equivalence class.
For every matrix $S_{n}$ of type $\mathbf{A}_n$, let $\tilde{S_{n}}$ be the corresponding binary matrix of type $\tilde{\mathbf{A}}_{n}$, i.e., the binary matrix with element $\tilde{s}_{ij}$ defined by
\begin{equation}
\tilde{s}_{ij}=\left\{\begin{array}{rcl}0,&&\mbox{if $s_{ij}=0$}\\1,&&\mbox{if $s_{ij}\ne 0$}\end{array}\right.
\end{equation}
Let every variable element of a binary matrix $\tilde{S}_{n}$ of type $\tilde{\mathbf{A}}_{n}$ have probability $r$ of being 1, and probability $1-r$ of being 0. Then let $\tilde{u}(\tilde{\mathbf{A}}_{n},\mathfrak{X})$ be the real number $\tilde{u}$ of least absolute value such that a matrix of type $\tilde{\mathbf{A}}_{n}$ can have determinant $\tilde{u}$ with non-zero probability if the variable elements of matrices of type $\mathbf{A}_n$ are chosen from $\mathfrak{X}$, and let $\tilde{\Omega}(\tilde{\mathbf{A}}_{n},\mathfrak{X})$ be the set of all matrices $\tilde{S}_{n}$ of type $\tilde{\mathbf{A}}_{n}$ such that $\det\tilde{S}_{n}=\tilde{u}(\tilde{\mathbf{A}}_{n},\mathfrak{X})$ with non-zero probability. In addition, for every $i\le i_{\max}$, let $\Omega_{i}(\mathbf{A}_n,\mathfrak{X})$ be the subset of $\Omega(\mathbf{A}_n,\mathfrak{X})$ that consists of matrices of type $\mathbf{A}_n$ that have $i$ variable elements with a non-zero value, and let $\tilde{\Omega}_{i}(\tilde{\mathbf{A}}_{n},\mathfrak{X})$ be the set of all matrices in $\tilde{\Omega}(\tilde{\mathbf{A}}_{n},\mathfrak{X})$ that have $i$ variable elements with a value of 1. ($i_{\max}$ is the maximum number of variable elements with a value of 1 that a matrix of type $\tilde{\mathbf{A}}_{n}$ which has determinant $\tilde{u}(\tilde{\mathbf{A}}_{n},\mathfrak{X})$ with non-zero probability can have.)
For type $\mathbf{B}_n$, define the following in a similar manner: $u(\mathbf{B}_n,\mathfrak{X}),\ \Omega(\mathbf{B}_n,\mathfrak{X})$, 
$\tilde{u}(\tilde{\mathbf{B}}_{n},\mathfrak{X}),\ \tilde{\Omega}(\tilde{\mathbf{B}}_{n},\mathfrak{X})$, and, for every $i\le i_{\max},\ \ \Omega_{i}(\mathbf{B}_n,\mathfrak{X})$ and $\tilde{\Omega}_{i}(\tilde{\mathbf{B}}_{n},\mathfrak{X})$. Then do likewise for type $\mathbf{C}_n$. 
For example, $u(\mathbf{C}_2,[0,2])=1$, and
\begin{equation}\label{EqC2CntOmega_p2}
\Omega(\mathbf{C}_{2},[0,2])=\left\{\left(\begin{array}{cc}1&0\\
0&1\\
\end{array}\right),\\
\left(\begin{array}{cc}1&x\\
0&1\\
\end{array}\right),\\
\left(\begin{array}{cc}1&0\\
y&1\\
\end{array}\right)\right\},
\end{equation}
where $x$ and $y$ are elements of $\mathfrak{X}-\{0\}$.  The probability that a matrix $C_{2}$ of type $\mathbf{C}_2$ is $\left(\begin{array}{cc}1&0\\0&1\end{array}\right)$ is $(1-r)^{2}$, since the probability that $c_{12}=0$ is $1-r$, as is the probability that $c_{21}=0$. The probability that $C_{2}$ is in the equivalence class of $\left(\begin{array}{cc}1&x\\0&1\end{array}\right)$ is $r\cdot(1-r)$, because the probability that $c_{12}\ne 0$ is $r$ and the probability that $c_{21}=0$ is $1-r$.  Similarly, the probability that $C_2$ is in the equivalence class of $\left(\begin{array}{cc}1&0\\y&1\end{array}\right)$ is $r(1-r)$.  
As an example of the discrete case, let $\mathfrak{X}=\{0,1/2,2\}$, and let $p(1/2)=r/2=p(2)$.  
Then a matrix $C_{2}$ of type $\mathbf{C}_2$ can have determinant 0 with non-zero probability (i.e., $u(\mathbf{C}_2,\{0,1/2,2\})=0$), because the probability that the term $c_{12}c_{21}$ in the expansion of $\det C_{2}$ via $\mathfrak{S}_{2}$ is equal to 1 is non-zero.  Note that
\begin{equation}\label{EqC2DisOmega_p2}
\Omega(\mathbf{C}_{2},\{0,1/2,2\})=\left\{\left(\begin{array}{cc}1&2\\
\displaystyle \frac{1}{2}&1\displaystyle \\
\end{array}\right),\\
\displaystyle \left(\begin{array}{cc}1&\displaystyle \frac{1}{2}\\
2&1\\
\end{array}\right)\right\}.
\end{equation}
The probability that $C_{2}=\displaystyle \left(\begin{array}{cc}1&2\displaystyle \\\frac{1}{2}&1\displaystyle \\\end{array}\right)$ is $(r/2)^{2}$, as is the probability that $C_{2}=\displaystyle \left(\begin{array}{cc}1&\displaystyle \frac{1}{2}\\2&1\displaystyle \\\end{array}\right)$. 
Thus
\begin{equation}\label{EqC2CntOmegaHat_p2}
\tilde{\Omega}(\tilde{\mathbf{C}}_{2},[0,2])=\left\{\left(\begin{array}{cc}1&0\\0&1\\
\end{array}\right),\\
\left(\begin{array}{cc}1&1\\0&1\\\end{array}\right),\\
\left(\begin{array}{cc}1&0\\1&1\\
\end{array}\right)\right\},
\end{equation}
and
\begin{equation}\label{EqC2DisOmegaHat_p2}
\tilde{\Omega}(\tilde{\mathbf{C}}_{2},\{0,1/2,2\})=\left\{\left(\begin{array}{cc}1&1\\1&1\\\end{array}\right)\right\}
\end{equation}
Partitioning the sets $\Omega(\mathbf{C}_{2},[0,2]),\ \Omega(\mathbf{C}_{2},\{0,1/2,2\}),\ \tilde{\Omega}(\tilde{\mathbf{C}}_{2},[0,2])$, and $\tilde{\Omega}(\tilde{\mathbf{C}}_{2},\{0,1/2,2\})$ according to the number $i$ of variable elements which are non-zero, we find that
\begin{equation}\label{EqC2CntEquivallentClass_p2}
\Omega_{0}(\mathbf{C}_{2},[0,2])=\left\{\left(\begin{array}{cc}1&0\\0&1\\\end{array}\right)\right\},\,\,\,\Omega_{1}(\mathbf{C}_{2},[0,2])=\left\{\left(\begin{array}{cc}1&x\\0&1\\\end{array}\right),\\
\left(\begin{array}{cc}1&0\\y&1\\\end{array}\right)\right\}
\end{equation}
\begin{equation}\label{EqC2CntHatEquivallentClass_p2}
\tilde{\Omega}_{0}(\tilde{\mathbf{C}}_{2},[0,2])=\left\{\left(\begin{array}{cc}1&0\\0&1\\\end{array}\right)\right\},\,\,\,\tilde{\Omega}_{1}(\tilde{\mathbf{C}}_{2},[0,2])=\left\{\left(\begin{array}{cc}1&1\\0&1\\\end{array}\right),\\
\left(\begin{array}{cc}1&0\\1&1\\\end{array}\right)\right\}
\end{equation}
and 
\begin{equation}\label{EqC2DisEquivallentClass_p2}
\Omega_{i}(\mathbf{C}_{2},\{0,1/2,2\})=\emptyset,\  i=0,1,\,\,\,\displaystyle \Omega_{2}(\mathbf{C}_{2},\{0,1/2,2\})=\left\{\left(\begin{array}{cc}1&2\displaystyle \\\frac{1}{2}&1\displaystyle \\
\end{array}\right),\\
\displaystyle \left(\begin{array}{cc}1&\displaystyle \frac{1}{2}\\2&1\displaystyle \\
\end{array}\right)\right\}
\end{equation}
\begin{equation}\label{EqC2DisHatEquivallentClass_p2}
\tilde{\Omega}_{i}(\tilde{\mathbf{C}}_{2},\{0,1/2,2\})=\emptyset,\  i=0,1,\,\,\,\tilde{\Omega}_{2}(\tilde{\mathbf{C}}_{2},\{0,1/2,2\})=\left\{\left(\begin{array}{cc}1&1\\1&1\\
\end{array}\right)\right\}
\end{equation}
For sets $S,\, T$, we will use $S\simeq T$ to denote that there is a one-to-one correspondence between $S$ and $T$.
\begin{prop}\label{Prop6_p2}
If $\mathfrak{X}$ is continuous, then
\begin{enumerate}\item[1.]  $u(\mathbf{A}_n,\mathfrak{X})=\tilde{u}(\tilde{\mathbf{A}}_{n},\mathfrak{X})$
\item[2.]  $\Omega_{i}(\mathbf{A}_{n},\mathfrak{X})\cong\tilde{\Omega}_{i}(\tilde{\mathbf{A}}_{n},\mathfrak{X})$ for every $i\le i_{\max}$
\item[3.]  the following are equivalent for a matrix $S_{n}$ of type $\mathbf{A}_n$:
\begin{enumerate}\item[(a)]  $\det S_{n}=u(\mathbf{A}_n,\mathfrak{X})$ with non-zero probability
\item[(b)]  $\mathrm{per}\, S_{n}=u(\mathbf{A}_n,\mathfrak{X})$ with non-zero probability
\item[(c)]  $\det\tilde{S}_{n}=\tilde{u}(\tilde{\mathbf{A}}_{n},\mathfrak{X})$ with non-zero probability
\item[(d)]  $\mathrm{per}\,\tilde{S}_{n}=\tilde{u}(\tilde{\mathbf{A}}_{n},\mathfrak{X})$ with non-zero probability
\end{enumerate}\end{enumerate}
The counterparts of all of these statements hold for types $\mathbf{B}_n$ and $\mathbf{C}_n$ as well. 
If $\mathfrak{X}$ is discrete, then statement 1 holds for types $\mathbf{A}_n$ and $\mathbf{B}_n$, but it does not hold in general for type $\mathbf{C}_n$; also, neither statement 2 nor statement 3 holds in general (for any of the three types).
\end{prop}
{\bf Proof}\ \ \ 
We have already established this for the case where $\mathfrak{X}$ is continuous: For types $\mathbf{A}_n$ and $\mathbf{B}_n$, we showed that the least value of $u$ such that $\det S_{n}=u$ with non-zero probability is 0, and that $\det S_{n}=0$ with non-zero probability if and only if $\mathrm{per}\,\tilde{S}_{n}=0$ with non-zero probability. For type $\mathbf{C}_n$, we showed that the least value of $u$ such that $\det S_{n}=u$ with non-zero probability is 1, and that $\det S_{n}=1$ with non-zero probability if and only if $\mathrm{per}\,\,\tilde{S}_{n}=1$. The truth of statement 2 is obvious. For type $\mathbf{A}_n$ or $\mathbf{B}_n$, the equivalence of 3(a) and 3(b)  follows from our proof that $\det S_{n}=0$ with non-zero probability if and only if every term in the expansion of $\det S_{n}$ via $\mathfrak{S}_{n}$ is 0, which is true if and only if every term in the expansion of $\mathrm{per}\, S_{n}$ via $\mathfrak{S}_{n}$ is 0. For type $\mathbf{C}_n$, the equivalence of 3(a) and 3(b) follows from our proof that $\det S_{n}=1$ with non-zero probability if and only if every non-diagonal term in the expansion of $\det S_{n}$ via $\mathfrak{S}_{n}$ is 0, which is true if and only if every non-diagonal term in the expansion of $\mathrm{per}\, S_{n}$ via $\mathfrak{S}_{n}$ is 0.
If $\mathfrak{X}$ is discrete, then for types $\mathbf{A}_n$ and $\mathbf{B}_n$, the least value of $u$ such that $\det S_{n}=u$ with non-zero probability is 0, as is the least value of $u$ for which $\det\tilde{S}_{n}=u$ with non-zero probability. Both of these facts are witnessed by matrices that have a single row or column comprised entirely of variable elements with a value of 0.
For a counterexample to statement 1 in type $\mathbf{C}_n$, let $\mathfrak{X}=\{0,1/2\}$. There are only four matrices of type $\mathbf{C}_2$ with variable elements chosen from $\mathfrak{X}$: \[\begin{array}{cccc}\left(\begin{array}{cc}1&0\\0&1\\
\end{array}\right)& \left(\begin{array}{cc}1&\frac{1}{2}\\0&1\\\end{array}\right)&\left(\begin{array}{cc}1&0\\\frac{1}{2}&1\\
\end{array}\right)& \left(\begin{array}{cc}1&\frac{1}{2}\\\frac{1}{2}&1\\
\end{array}\right)\end{array}\] Thus $u(\mathbf{C}_2,\mathfrak{X})=3/4$, as witnessed by $S_2=\left(\begin{array}{cc}1&\frac{1}{2}\\
\frac{1}{2}&1\\
\end{array}\right)$, which has determinant 3/4 with non-zero probability. The associated binary matrix is $\tilde{S}_2=\left(\begin{array}{cc}1&1\\1&1\\
\end{array}\right)$, which has determinant 0 with non-zero probability, so $\tilde{u}(\tilde{\mathbf{C}}_2,\mathfrak{X})=0$.
For a counterexample to statement 2 in the discrete case, let $\mathfrak{X}=\{0,1/2,1,2\}$. If $S_{3}=\displaystyle \left(\begin{array}{ccc}1&0&\displaystyle \frac{1}{2}\\0&1&0\displaystyle \\2&1&1\displaystyle \\\end{array}\right)$, then $S_{3}$ can be viewed as a matrix of type $\mathbf{A}_3$, type $\mathbf{B}_3$, or type $\mathbf{C}_3$. The associated binary matrix is $\tilde{S}_{3}=\left(\begin{array}{ccc}1&0&1\\0&1&0\\1&1&1\\\end{array}\right)$. Since $\{0,1\}\subset \mathfrak{X}$, the $3\times 3$ matrix $T_{3}$ whose elements are equal to those of $\tilde{S}_{3}$ can also be viewed as a matrix of type $\mathbf{A}_3$, type $\mathbf{B}_3$, or type $\mathbf{C}_3$. Regardless of type, matrices $S_{3},\ T_{3}$, and $\tilde{S}_{3}$ all have determinant 0 with non-zero probability, so $u$ and $\tilde{u}$ are 0 (even for type $\mathbf{C}_3$).  Thus $S_{3}$ and $T_{3}$ serve as witnesses to the failure of statement 2: The fact that $\tilde{S}_{3}$ is the associated binary matrix for both $S_{3}$ and $T_{3}$ means that $\Omega_{6}(\mathbf{A}_3,\mathfrak{X})\not\simeq\tilde{\Omega}_{6}(\tilde{\mathbf{A}}_{3},\mathfrak{X}),\ \ \Omega_{4}(\mathbf{B}_3,\mathfrak{X})\not\simeq\tilde{\Omega}_{4}(\tilde{\mathbf{B}}_{3},\mathfrak{X})$, and $\Omega_{3}(\mathbf{C}_3,\mathfrak{X})\not\simeq\tilde{\Omega}_{3}(\tilde{\mathbf{C}}_{3},\mathfrak{X})$.  Also, $\mathrm{per}\,S_{3}=2=\mathrm{per}\,\tilde{S}_{3}$. Thus for all three types of matrices, $S_3$ also serves as a counterexample to statement 3, though not as a counterexample to the equivalence of 3(a) and 3(c), and not as a counterexample to the equivalence of 3(b) and 3(d). 
For a counterexample to the equivalence of 3(a) and 3(c) (for all three types of matrices) in the discrete case, let $\mathfrak{X}=\{0,1,2\}$. Then $u=0=\tilde{u}$ for types $\mathbf{A}_3,\ \mathbf{B}_3$, and $\mathbf{C}_3$, as witnessed by the $3\times 3$ matrix $M=\left(\begin{array}{ccc}1&1&0\\1&1&0\\0&1&1\\\end{array}\right)$, which has determinant 0 with non-zero probability and can be viewed as a matrix of type $\mathbf{A}_3$, type $\mathbf{B}_3$, or type $\mathbf{C}_3$. Moreover, since $\{0,1\}\subset\mathfrak{X}$, the elements of the associated binary matrix of $M$ are identical to those of $M$ itself, so $\det\tilde{M}=0$ with non-zero probability. If $S_{3}=\left(\begin{array}{ccc}1&0&1\\1&1&0\\2&1&1\\\end{array}\right)$, then $S_{3}$ can also be viewed as a matrix of type $\mathbf{A}_3$, type $\mathbf{B}_3$, or type $\mathbf{C}_3$. The associated binary matrix is $\tilde{S}_{3}=\left(\begin{array}{ccc}1&0&1\\1&1&0\\1&1&1\\\end{array}\right)$.  
Thus $\det S_{3}=0$ and $\det\tilde{S}_{3}=1$, which shows that 3(a) and 3(c) are not equivalent in general.  Hence 3(b) and 3(d) are the only two items in statement 3 whose equivalence we have neither proved nor refuted (in the case where $\mathfrak{X}$ is discrete). $\blacksquare$
\begin{prop}\label{prop7_p2}
Let $\mathfrak{X}^{(\mathrm{cnt})}$ and $\mathfrak{X}^{(\mathrm{dis})}$ be a continuous set and a discrete set, respectively, such that $0\in \mathfrak{X}^{(\mathrm{dis})}\subseteq \mathfrak{X}^{(\mathrm{c}\mathrm{n}\mathrm{t})}$. Then $\tilde{\Omega}(\tilde{\mathbf{A}}_{n},\mathfrak{X}^{(\mathrm{dis})})\sqsupseteq\tilde{\Omega}(\tilde{\mathbf{A}}_{n},\mathfrak{X}^{(\mathrm{c}\mathrm{n}\mathrm{t})})$ and $\tilde{\Omega}(\tilde{\mathbf{B}}_{n},\mathfrak{X}^{(\mathrm{dis})})\sqsupseteq\tilde{\Omega}(\tilde{\mathbf{B}}_{n},\mathfrak{X}^{(\mathrm{c}\mathrm{n}\mathrm{t})})$, where $\subseteq$ and $\sqsubseteq$ denote set inclusion for sets of numbers and sets of matrices, respectively.  However, there exist $\mathfrak{X}^{(\mathrm{cnt})}$ and $\mathfrak{X}^{(\mathrm{dis})}$ such that $0\in \mathfrak{X}^{(\mathrm{dis})}\subseteq \mathfrak{X}^{(\mathrm{c}\mathrm{n}\mathrm{t})}$ but $\tilde{\Omega}(\tilde{\mathbf{C}}_{n},\mathfrak{X}^{(\mathrm{dis})})\not\sqsupseteq\tilde{\Omega}(\tilde{\mathbf{C}}_{n},\mathfrak{X}^{(\mathrm{c}\mathrm{n}\mathrm{t})})$.
\end{prop}
{\bf Proof}\ \ \ The first statement follows from the fact that every element of $\tilde{\Omega}(\tilde{\mathbf{A}}_{n},\mathfrak{X}^{(\mathrm{cnt})})$ is a matrix with at least one row or column comprised entirely of variable elements with a value of 0, as is every element of $\tilde{\Omega}(\tilde{\mathbf{B}}_{n},\mathfrak{X}^{(\mathrm{cnt})})$. For example, let $\mathfrak{X}^{(\mathrm{cnt})}=[0,2]$ and $\mathfrak{X}^{(\mathrm{dis})}=\{0,1/2,2\}$. Then $u(\mathbf{A}_{2},\mathfrak{X}^{(\mathrm{cnt})})=0=u(\mathbf{A}_{2},\mathfrak{X}^{(\mathrm{d}\mathrm{i}\mathrm{s})})$ (therefore $\tilde{u}(\tilde{\mathbf{A}}_{2},\mathfrak{X}^{(\mathrm{cnt})})=0=\tilde{u}(\tilde{\mathbf{A}}_{2},\mathfrak{X}^{(\mathrm{d}\mathrm{i}\mathrm{s})})$), so
\begin{equation}
\nonumber\tilde{\Omega}(\tilde{\mathbf{A}}_{2},\mathfrak{X}^{(\mathrm{cnt})})=\left\{\left(\begin{array}{cc}0&0\\
0&0\\
\end{array}\right),\\
\left(\begin{array}{cc}1&0\\
0&0\\
\end{array}\right),\\
\left(\begin{array}{cc}0&1\\
0&0\\
\end{array}\right),\\
\left(\begin{array}{cc}0&0\\
1&0\\
\end{array}\right),\\
\left(\begin{array}{cc}0&0\\
0&1\\
\end{array}\right),\right.
\end{equation}
\begin{equation}\label{EqA2CntOmegaHat_p2}
\left.\hspace*{.5in}\left(\begin{array}{cc}1&1\\
0&0\\
\end{array}\right),\\
\left(\begin{array}{cc}0&0\\
1&1\\
\end{array}\right),\\
\left(\begin{array}{cc}1&0\\
1&0\\
\end{array}\right)\\
\left(\begin{array}{cc}0&1\\
0&1\\
\end{array}\right)\right\},
\end{equation}
\begin{equation}\label{EqA2DisOmegaHat_p2}
\tilde{\Omega}(\tilde{\mathbf{A}}_{2},\mathfrak{X}^{(\mathrm{dis})})=\tilde{\Omega}(\tilde{\mathbf{A}}_{2},\mathfrak{X}^{(\mathrm{c}\mathrm{n}\mathrm{t})})\bigcup\left\{\left(\begin{array}{cc}1&1\\1&1\\\end{array}\right)\right\}
\end{equation}
The example given in (\ref{EqC2CntOmegaHat_p2}) and (\ref{EqC2DisOmegaHat_p2}) illustrates the truth of the second statement.  $\blacksquare$
We can exploit the difference between the continuous and discrete cases to create an interesting example in which the real number $u$ of least absolute value such that a matrix of type $\mathbf{C}_2$ can have determinant $u$ with non-zero probability in the continuous case differs from its counterpart in the discrete case.
Let $\mathfrak{X}^{(\mathrm{cnt})}=[0,1]$ and $\mathfrak{X}^{(\mathrm{dis})}=\{0,1\}$.  Then $u(\mathbf{C}_{2},\mathfrak{X}^{(\mathrm{cnt})})=1$ (therefore $\tilde{u}(\tilde{\mathbf{C}}_{2},\mathfrak{X}^{(\mathrm{cnt})})=1$), and
\begin{equation}\nonumber
\Omega(\mathbf{C}_{2},\mathfrak{X}^{(\mathrm{cnt})})=\left\{\left(\begin{array}{cc}1&0\\0&1\\\end{array}\right),\\
\left(\begin{array}{cc}1&x\\0&1\\\end{array}\right),\\
\left(\begin{array}{cc}1&0\\y&1\\\end{array}\right)\right\}\,\,\,
\end{equation}
\begin{equation}\nonumber
\tilde{\Omega}(\tilde{\mathbf{C}}_{2},\mathfrak{X}^{(\mathrm{cnt})})=\left\{\left(\begin{array}{cc}1&0\\0&1\\\end{array}\right),\\
\left(\begin{array}{cc}1&1\\0&1\\\end{array}\right),\\
\left(\begin{array}{cc}1&0\\1&1\\
\end{array}\right)\right\},
\end{equation}
where $x$ and $y$ are elements of $\mathfrak{X}^{(\mathrm{cnt})}-\{0\}$.
Also, $u(\mathbf{C}_{2},\mathfrak{X}^{(\mathrm{dis})})=0$ (therefore $\tilde{u}(\tilde{\mathbf{C}}_{2},\mathfrak{X}^{(\mathrm{dis})})=0$), and
\begin{equation}\label{EqC2DisOmegaAndOmegaHat_p2}
\Omega(\mathbf{C}_{2},\mathfrak{X}^{(\mathrm{dis})})=\left\{\left(\begin{array}{cc}1&1\\1&1\\\end{array}\right)\right\}=\tilde{\Omega}(\tilde{\mathbf{C}}_{2},\mathfrak{X}^{(\mathrm{d}\mathrm{i}\mathrm{s})})
\end{equation}
Thus
\begin{equation}\nonumber
\tilde{\Omega}(\tilde{\mathbf{C}}_{2},\mathfrak{X}^{(\mathrm{cnt})})\cap\tilde{\Omega}(\tilde{\mathbf{C}}_{2},\mathfrak{X}^{(\mathrm{d}\mathrm{i}\mathrm{s})})=\emptyset.
\end{equation}
The probability that $\det C_{2}=1$ ($\mathrm{per}C_{2}=1$) in the continuous case is 
\begin{equation}\nonumber
P_{\det C_{2}=1}^{(\mathrm{cnt})}(r)=1r^{0}(1-r)^{2}+2r(1-r)=1-r^{2},
\end{equation}
while in the discrete case the probability that $\det C_{2}=0$ is
\begin{equation}\nonumber
P_{\det C_{2}=0}^{(\mathrm{dis})}(r)=1r^{2}(1-r)^{0}=r^{2}
\end{equation}
Therefore,
\begin{equation}\label{EQEventBifurcation_p2}
P_{\det C_{2}=1}^{(\mathrm{cnt})}(r)+P_{\det C_{2}=0}^{(\mathrm{d}\mathrm{i}\mathrm{s})}(r)=1
\end{equation}


\begin{thebibliography}{99}
\bibitem{BenAmirHalevi}    Ben-Dor, A., Halevi, S. Zero-one Permanent is $\# P$
-complete, a simpler proof. {\it Proceedings of the 2nd Israel Symposium on the Theory and Computing Systems} 1993; 108--117.
\bibitem{EverettStein}    Everett, C.J., Stein, P.R. The asymptotic number of (0,1)-matrices with zero Permanent. 
{\it Discrete Math}. 1973; {\bf 6}: 29--34.
\bibitem{Gessel} Gessel, I.M. Counting acyclic digraphs by sources and sinks. {\it Discrete Math}. 1996; {\bf 160}: 253--258.
\bibitem{Marcus} Marcus, M. {\it Permanents, Encyclopedia of Mathematics and Its Applications Vol. 6}. Addison-Wesley, 1978.
\bibitem{Sloane} Sloane, N.J.A. {\it The On-Line Encyclopedia of Integer Sequences} (published electronically at http://www.research.att.com/$_{\widetilde{\hphantom{mx}}}$njas/sequences/). 1996--2007.
\bibitem{Valiant} Valiant, L.G. The Complexity of Computing the Permanent. {\it Theoretical Computer Science}. 1979; {\bf 8}: 189--201.
\end{thebibliography}
\end{document}